\documentclass[preprint2]{aastex}

\shorttitle{Detailed chemical abundance of NGC 1851}
\shortauthors{Villanova et al.}

\begin{document}

\title{Detailed abundances of Red Giants in the Globular Cluster NGC~1851:
       C+N+O and the Origin of Multiple Populations}

\author{S. Villanova and D. Geisler}
\affil{Departamento de Astronomia, Casilla 160, 
       Universidad de Concepcion, Chile}
\email{svillanova@astro-udec.cl}

\and

\author{G. Piotto}
\affil{Dipartimento di Astronomia, Universit\`a di Padova, Vicolo 
       dell'Osservatorio 3, I-35122 Padua, Italy}

\begin{abstract}
We present chemical abundance analysis of a sample of 15 red giant branch (RGB)
stars of the Globular Cluster NGC~1851 distributed along the two RGBs of the (v, v-y) CMD.
We determined abundances for C+N+O, Na, $\alpha$, iron-peak, and s-elements.
We found that the two RGB populations significantly differ in their
light (N,O,Na) and s-element content. On the other hand, they
do not show any significant difference in their $\alpha$ and iron-peak
element content. More importantly, the two RGB populations do not 
show any significant difference in their total C+N+O content. 

Our results do not support previous hypotheses suggesting that the origin of the two 
RGBs and the two subgiant branches of the cluster is related to a different content of
either $\alpha$ (including Ca) or iron-peak 
elements, or C+N+O abundance, due to a second generation polluted
by SNeII.
\end{abstract}

\keywords{globular clusters: general --- globular clusters: individual(NGC 1851)}

\section{Introduction}

Recent investigations have fostered new interest in globular cluster (GC) stellar
populations. The traditional vision of GCs hosting simple, single age, single
metallicity stars has been shattered by the discovery of multiple sequences
in the color-magnitude diagrams (CMD) and of large dispersions in light or heavy
element abundances, often showing well defined patterns, which are the result
of nuclear processes in a previous generation of stars. \\
NGC 1851 is one of the GCs where multiple stellar populations have been clearly
identified.
Indeed, in recent years, this GC has been  the subject 
of a wide photometric and spectroscopic observational campaign. 
First of all,  \citet{mil08} have identified a double subgiant branch (SGB) in the cluster
CMD, clearly suggesting the presence of at least two stellar generations. 
As discussed by \citet{mil08}, if the SGB split would be due to a different age
between the two populations with similar metal content, 
the second generation of stars should have formed
$\sim$ 1 Gyr after the first one. However other factors could be at work.
As pointed out by \citet{cas08} and \citet{ven09}, 
the SGB splitting could also be explained by a difference in the total C+N+O content, 
with an abundance for the fainter SGB 2-3 times larger than that of the brighter one.
In this case, the age spread would be smaller, of the order of a few hundred Myrs.

Unfortunately, we have scarce information on detailed abundances of
NGC 1851 stars. \citet{hes82} collected low
resolution spectra for 18 stars.  They found a constant
iron content, but 3 out of 8 of their bright RGB stars show
anomalously strong CN bands.
More recently, \citet{yon08} analyzed UVES spectra
of eight giants. With all the limitations coming from the small number
statistics,  \citet{yon08} results seem to indicate the presence
of a Na-O anticorrelation, and the presence of a large star to star
abundance scatter of s-process elements Zr and La, with some hints
of a bimodal distribution. They also find a large star to star
variation in the strength of the Ba lines. They also show that the split 
of the RGB found by \citet{cal07} in a Str\"omgren photometry CMD
is related to the double SGB.  
Interestingly, the fraction of La/Zr-strong
stars (40\%) is tantalizing similar to the fraction of SGB-faint stars
(45\%),  and to the fraction of the CN-strong,
Ba-strong, and Sr-strong RGB stars (40\%) of \citet{hes82}.
But the sample statistics for stars with abundance estimates is
very poor.\\
Recently, \citet{yon09} suggested the presence of 
a large C+N+O spread ($\sim$0.6 dex) among NGC 1851 giant stars. Their work is based on high
resolution spectra of four stars. This spread could imply that
one population has 4 times larger C+N+O content with respect to the other,
supporting the scenario proposed by \citet{cas08} and \citet{ven09}. However, also in
this case, the statistics is very poor.

\begin{figure}[!ht]
\includegraphics[angle=0,scale=.40]{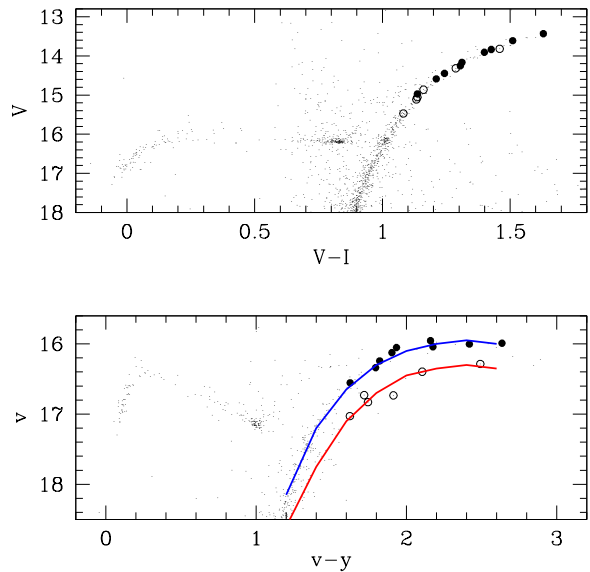}
\caption{V vs. V-I and v vs. v-y CMDs of our target stars.
The double RGB is visible in the v vs. v-y CMD but not in
the V vs. V-I CMD. The blue RGB
is indicated as a blue line, the red RGB as a red line.
Ba-poor stars are plotted as filled circles, while Ba-rich as 
open circles.}
\label{f1}
\end{figure}

In order to account for such observational results, it 
has been proposed that a second generation of stars may have
formed from a residual reservoir of gas polluted by intermediate
mass asymptotic giant branch (AGB) star ejecta \citep{dan02}, or from the ejecta
of massive fast rotating stars \citep{dec07}.\\
A very recent work by \citet{lee09} proposes another
possibility. Using Str\"omgren photometry and the Ca narrow-band filter centered on
the H-K Ca lines at $\sim$3950 \AA, \citet{lee09} find a significant color
split in the RGB of many clusters, including NGC~1851.
Because of the observed split, \citet{lee09} propose that the two RGBs have a different Ca content, 
of the order of $0.3$ dex.
In this case, the second generation of stars could originate from gas polluted by SNeII
explosions, the only mechanism able to produce a Ca enhancement.
However, this scenario is quite controversial, because multiple CMD sequences have
been found in other clusters, where abundances based on high resolution
spectra give a negligible or very low Ca spread. 
For example, there is the case of M4, where \citet{mar08} and
\citet{car10} found no Ca spread at all, while \citet{lee09} presented a CMD with a
double RGB.
In M4, a clear RGB spread is visible in ultraviolet broad-band photometry (see
Fig. 11 of \citet{mar08}), but, as discussed in \citet{mar08}, the photometric
spread can be better explaind by a spread in C,N,O.\\

The only way to solve the enigma is to measure 
C,N,O, $\alpha$, iron-peak, and s-process element abundances for
a statistically significant sample of RGB stars.
For this purpose we observed a sample
of 15 RGB stars distributed among the two RGBs of the Str\"omgren 
(v,v-y) CMD with high and medium resolution spectroscopy. The observations,
data reduction, chemical abundance analysis, and results will be presented in
the following sections. 

In this letter we discuss only the results on Fe, Ca, s-process, and C+N+O content,
i.e. the most relevant elements in the context of the present debate on multiple
stellar populations in GCs. Full details will be presented in a subsequent paper.

\begin{figure}[!h]
\includegraphics[angle=0,scale=.40]{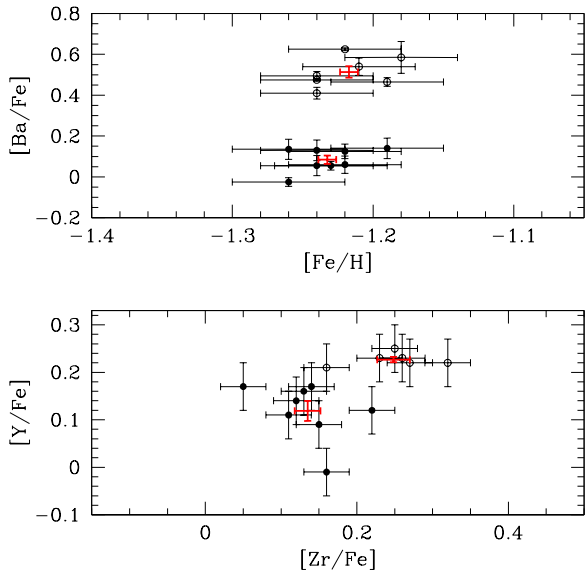}
\caption{[Ba/Fe] vs. [Fe/H] and [Y/Fe] vs. [Zr/Fe] for our target
stars. Ba-poor stars are plotted as filled circles, while Ba-rich as open
circles. Red crosses give mean values and errors of the mean for the two
groups of stars.}
\label{f2}
\end{figure}

\section{Observations}

Our dataset consists of high and medium resolution spectra collected in November-December 2008.
The spectra come from single $\sim$1500 and $\sim$500 s exposures, 
obtained with the MIKE and Mage slit spectrographs \footnote{See www.lco.cl} respectively,
mounted at the Magellan/Clay telescope.
Weather conditions were excellent with a typical seeing of 0.5 arcsec. 
We selected 15 isolated stars from 0.5 magnitudes above the RGB-bump to
the tip of the RGB, in the magnitude range 15.5$<V<$13.5.
The MIKE spectra have a spectral coverage from 4900 to 8700 \AA\ with a
resolution of $\sim$ 32000 and S/N$>$100 at 6300 \AA.
The Mage spectra have a spectral coverage from 3700 to 9500 \AA\ with a
resolution of $\sim$ 5000 and S/N$>$50 at 4300 \AA.

Data were reduced using IRAF \footnote{IRAF is distributed by the National
Optical Astronomy Observatory, which is operated by the Association of
Universities for Research in Astronomy, Inc., under cooperative agreement
with the National Science Foundation} including bias
subtraction, flat-field correction, wavelength calibration, scattered-light
and sky subtraction, and spectral rectification.\\
In the case of MIKE observations, spectra were corrected for fringing using the
spectrum of a hot fast rotating star, which gave the best result compared with
other methods. However, fringing did not affect the spectral region we were
interested in (below 8100 \AA). 

Targets were carefully selected in order to have stars in each of the two
distinct RGBs clearly separated in the high quality 
v vs. v-y Str\"omgren photometry CMD (see Fig.~\ref{f1}, lower 
panel) kindly provided by F.\ Grundahl. In the following, we used also VI Cousins
photometry from Y.\ Momany. Note that there is no RGB split 
in the V vs. V-I CMD (Fig.~\ref{f1}, upper panel). 
Filled and open circles in Fig.~\ref{f1} are Ba-poor and Ba-rich stars, respectively (as 
measured from our spectra, see Sec. 4). All stars were observed with both
spectrographs.

Radial velocities were measured by the {\it fxcor} package in IRAF,
using a synthetic spectrum as a template. 
 The mean heliocentric value for our targets is 320.3$\pm$1.1 km/s, while
the dispersion is 4.4$\pm$0.8 km/s. \citet{har96} gives
320.5$\pm$0.6 km/s as heliocentric radial velocity for NGC1851, 
and the typical dispersion for a cluster of its mass is $\sim$4-5 km/s \citep{pry93}.
All our targets have a radial velocity within $\pm$15 km/s of the cluster radial velocity.
On the basis of this result, we conclude that all our targets 
are cluster members.

\section{Abundance analysis}

The chemical abundances for Na, Si, Ca, Ti, Fe, and Ni were
obtained from the equivalent widths (EWs) of the spectral lines. 
See \citet{mar08} for a more detailed explanation of the method we
used to measure the EWs and for the adopted solar abundances.
For the other elements (C, N, O, Y, Zr, Ba), whose lines are affected by
blending, we used the spectral synthesis method.  
Na presents few features in the spectrum, so in this case abundances derived from
the EWs were cross-checked with the spectral synthesis method in order to
obtain more accurate measurements.
MIKE spectra were used to obtain abundances for all the elements but C, which
was obtained from Mage data. Only lines not contaminated by 
telluric lines were used. 

\begin{figure}[!h]
\includegraphics[angle=0,scale=.40]{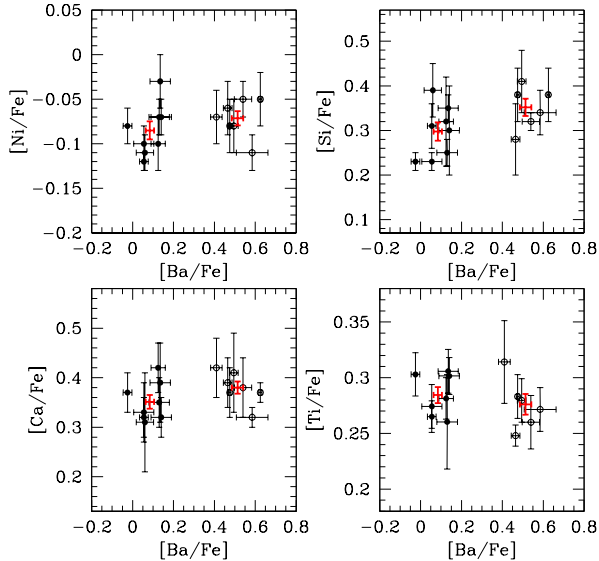}
\caption{Ni and $\alpha$-element abundances of our target
stars. Ba-poor stars are plotted as filled circles, while Ba-rich as open circles.
Red crosses are mean values and errors of the mean for the two
groups of stars.}
\label{f3}
\end{figure}

Atmospheric parameters were obtained in the following way. 
First of all T$_{\rm eff}$ was derived from the V-I color using the relation by 
\citet{alo99} and the reddening from \citet{har96}. 
Surface gravities (${\log(g)}$) were obtained from the canonical equation:

\begin{displaymath}
\log(\frac{g}{g_{\odot}}) = \log(\frac{M}{M_{\odot}}) + 4\cdot
\log(\frac{T_{\rm eff}}{T_{\odot}}) - \log(\frac{L}{L_{\odot}})
\end{displaymath}

where the mass ${M/M_{\odot}}$ was derived from 
isochrone fitting using the Padova database \citep{mari08}, and the
luminosity ${L/L_{\odot}}$ was obtained from the absolute magnitude ${M_{\rm V}}$
assuming an apparent distance modulus of ${(m-M)_{\rm V}}$=15.47 \citep{har96}. The
bolometric correction (BC) was derived by adopting the relation 
BC-T$_{\rm eff}$ from \citet{alo99}.
Finally, microturbulence velocity (${v_{\rm t}}$) was obtained from the
relation of \citet{gra96}.\\
These atmospheric parameters were considered as initial estimates and were refined during the
abundance analysis. As a first step atmospheric models were calculated using ATLAS9 \citep{kur70}
and assuming the initial estimate of T$_{\rm eff}$, ${\log(g)}$,
and ${v_{\rm t}}$, and the [Fe/H] value from \citet{har96}.\\ 
Then T$_{\rm eff}$, ${v_{\rm t}}$, and ${\log(g)}$ were adjusted and new
atmospheric models calculated in an interactive way in order to remove trends in
Excitation Potential (E.P.) and equivalent widths vs. abundance for 
${T_{\rm eff}}$ and ${v_{\rm t}}$ respectively, and to satisfy the ionization
equilibrium for ${\log(g)}$. FeI and FeII were used for this purpose. 
The [Fe/H] value of the model was changed at each iteration according to the
output of the abundance analysis.\\
The Local Thermodynamic Equilibrium (LTE) program MOOG \citep{sne73} was used
for the abundance analysis.

We checked the reliability of our atmospheric parameters in the following
way. Intrinsic V-I colours of our stars were obtained from our T$_{\rm eff}$
by inverting the \citet{alo99} equation. Comparing 
intrinsic and measured colours we
got a reddening of E(B-V)=0.01$\pm$0.01 (assuming E(V-I)=1.24$\times$E(B-V)).\\
This value is in very good agreement with \citet{har96}, that gives
E(B-V)=0.02.\\ 
A further check to test the reliability of our T$_{\rm eff}$ and log(g) values was
performed on Arcturus, which is an important reference for every study on RGB
stars.\\
In order to have statistically independent estimations,
we measured the atmospheric parameters on 3 high-resolution high-S/N spectra
obtained with HARPS, the high precision spectrograph mounted at the 3.6m
telescope in La Silla.\\ 
We obtained \ T$_{\rm eff}$=4290$\pm$11 K \ and
log(g)=\\2.00$\pm$0.02 as mean values and internal errors, which agree
extremely well with  literature values (T$_{\rm eff}$=4290$\pm$30 K,
log(g)=1.9$\pm$0.1, \citealt{gri99}).\\
Our conclusion is that our T$_{\rm eff}$ and log(g) scales appear free from
significant systematic errors.

The linelists for the chemical analysis were obtained from many sources
(\citet{gra03}, VALD \& NIST\footnote{See
  http://vald.astro.univie.ac.at/$\sim$vald/php/vald.php and http://physics.nist.gov/PhysRefData/ASD/lines\_form.html}, \citet{mcw94},
\citet{mcw98}, SPECTRUM\footnote{See   http://www.phys.appstate.edu/spectrum/spectrum.html 
and references therein}, and SCAN\footnote{See http://www.astro.ku.dk/$\sim$uffegj/}), 
and calibrated using the Solar-inverse technique by the spectral synthesis
method. For this purpose we used the high resolution, high S/N NOAO Solar spectrum
\citep{kur84}. 
For Ba lines we took the hyperfine splitting into account.
We emphasize the fact that all the linelists were calibrated on the Sun,
including those used for the spectral synthesis.\\
In particular, our determinations of 
C,N,O abundances are based on the G-band at 4310 \AA, the CN lines at 8003 \AA, and the
forbidden O line at 6300 \AA\ respectively. CN lines at 8003 \AA\ were used also
to estimate the C$^{12}$/C$^{13}$ ratio.\\
For the Sun we have log$\epsilon$(C)=8.49,\ log$\epsilon$(N)=7.95, and
log$\epsilon$(O)=8.83, values that agree with \citet{gre98} within 0.03 dex.\\
These three features were also checked on the high resolution, high S/N
spectrum of Arcturus by \citet{hin03} in order to verify if further
blends affecting cool RGB stars but not the Sun were present. We found no
significant contamination.\\
We obtained the following results for Arcturus: [C/Fe]=-0.10$\pm$0.05, 
\ [N/H]=+0.34$\pm$\\0.05, [O/Fe]=+0.38$\pm$0.05 (internal errors). These values well agree
with \citet{pet93}, which gives [C/Fe]=+0.0$\pm$0.1, 
[N/H]=+0.3$\pm$0.1, [O/Fe]=+0.4$\pm$0.1. We consider this as a 
test which proves the reliability of our linelists for C,N,O not only on
Sun-like stars, but also on cool RGB stars.\\
Abundances for C, N, and O were determined all together in an
interactive way, because for the temperature of our 
stars, carbon, nitrogen, and oxygen form molecules and as a consequence their
abundances are related to each other.\\ 
Our C$^{12}$/C$^{13}$ estimation (C$^{12}$/C$^{13}$$<$15 for all our targets) confirm that the 
stars are affected by evolutionary mixing, as expected from their position in
the CMD. This can affect the primordial C,N,O abundances separately, but not the total
C+N+O content because these elements are transformed one into the other during the
CNO cycle.

\begin{figure}[!ht]
\includegraphics[angle=0,scale=.40]{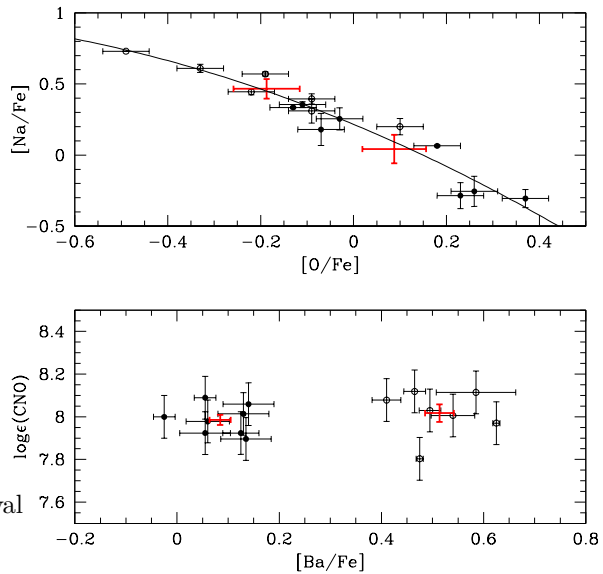}
\caption{Na-O anticorrelation and CNO abundances for our targets. 
Ba-poor stars are plotted as filled circles, while Ba-rich as open circles.
Red crosses are mean values and errors of the mean for the two
groups of stars.}
\label{f4}
\end{figure}

\section{Results and Discussion}

First of all, in Fig.~\ref{f2} we plot the abundance of Ba vs. Fe. (upper panel).
In this plot we clearly see that the Ba distribution is bimodal, while
the iron abundance for the two Ba groups is the same within the errors. 
In the following analysis, we divided our stars into two groups, one Ba-poor 
(filled circles in the figures), and the other Ba-rich (open circles).\\
The individual and mean abundances we obtained for the two groups are
summarized in Tab.~\ref{t1}.
For the mean abundances we reported also the internal errors (standard
deviation of the mean).

\begin{deluxetable}{lccccccccccc}
\tablecolumns{12}
\tablewidth{0pc}
\tablecaption{Abundances of the two groups of stars. Mean abundaces of the two
populations and the related internal errors are reported at the bottom of the table. Errors associated with
the measurements are plotted in Figs.~\ref{f2},\ref{f3}, and \ref{f4}.}
\tablehead{
\colhead{\small{Id}} & \colhead{\small{[Fe/H]}} & \colhead{\small{[Ba/Fe]}} & \colhead{\small{[Zr/Fe]}} &
\colhead{\small{[Y/Fe]}} & \colhead{\small{[Ni/Fe]}} & \colhead{\small{[Si/Fe]}} & \colhead{\small{[Ca/Fe]}} &
\colhead{\small{[Ti/Fe]}} & \colhead{\small{[Na/Fe]}} &  \colhead{\small{[O/Fe]}} & \colhead{\small{C+N+O}}
}
\startdata
\multicolumn{12}{c}{Ba-poor stars}\\
\hline
13 & -1.24 &  0.06 & 0.16 & -0.01 & -0.10 & 0.23 & 0.33 & 0.27 & -0.26 &  0.26 & 7.92\\ 
30 & -1.23 &  0.06 & 0.05 &  0.17 & -0.12 & 0.31 & 0.32 & 0.26 &  0.06 &  0.18 & 8.09\\ 
31 & -1.26 & -0.02 & 0.11 &  0.11 & -0.08 & 0.23 & 0.37 & 0.30 & -0.31 &  0.37 & 8.00\\ 
43 & -1.22 &  0.06 & 0.14 &  0.17 & -0.11 & 0.39 & 0.31 &  -   &  0.34 & -0.13 & 7.98\\ 
51 & -1.19 &  0.14 & 0.22 &  0.12 & -0.07 & 0.30 & 0.32 & 0.30 &  0.26 & -0.03 & 8.06\\ 
53 & -1.22 &  0.13 & 0.15 &  0.09 & -0.10 & 0.32 & 0.42 & 0.28 & -0.28 &  0.23 & 7.92\\ 
63 & -1.26 &  0.14 & 0.12 &  0.14 & -0.03 & 0.35 & 0.39 & 0.31 &  0.18 & -0.07 & 7.90\\ 
68 & -1.24 &  0.13 & 0.13 &  0.16 & -0.07 & 0.25 & 0.35 & 0.26 &  0.36 & -0.11 & 8.01\\ 
\hline
\multicolumn{12}{c}{Ba-rich stars}\\
\hline
35 & -1.24 &  0.48 & 0.26 &  0.23 & -0.08 & 0.38 & 0.37 & 0.28 &  0.31 & -0.09 & 7.80\\ 
14 & -1.21 &  0.54 & 0.32 &  0.22 & -0.05 & 0.32 & 0.38 & 0.26 &  0.20 &  0.10 & 8.01\\ 
16 & -1.24 &  0.50 & 0.16 &  0.21 & -0.08 & 0.41 & 0.41 & 0.28 &  0.40 & -0.09 & 8.03\\ 
18 & -1.19 &  0.47 & 0.25 &  0.25 & -0.06 & 0.28 & 0.39 & 0.25 &  0.73 & -0.49 & 8.12\\ 
20 & -1.24 &  0.41 & 0.23 &  0.23 & -0.07 &  -   & 0.42 & 0.31 &  0.57 & -0.19 & 8.08\\ 
8  & -1.18 &  0.59 & 0.27 &  0.22 & -0.11 & 0.34 & 0.32 & 0.27 &  0.44 & -0.22 & 8.11\\
9  & -1.22 &  0.63 &  -   &   -   & -0.05 & 0.38 & 0.37 &  -   &  0.61 & -0.33 & 7.97\\ 
\hline
\multicolumn{12}{c}{Mean abundances and internal errors}\\
\hline\hline
\small{Ba-poor} & \small{-1.23} & \small{+0.09} & \small{+0.14} & \small{+0.12} & \small{-0.09} & \small{+0.30} & \small{+0.35} & \small{+0.28} & \small{+0.04} & \small{+0.09} & \small{7.99}\\
\small{       } & \small{$\pm$0.01} & \small{$\pm$0.02} & \small{$\pm$0.02} & \small{$\pm$0.02} & \small{$\pm$0.01} & \small{$\pm$0.02} & \small{$\pm$0.01} & \small{$\pm$0.01} & \small{$\pm$0.10} & \small{$\pm$0.07} & \small{$\pm$0.02}\\
\hline
\small{Ba-rich} & \small{-1.22} & \small{+0.52} & \small{+0.25} & \small{+0.23} & \small{-0.07} & \small{+0.35} & \small{+0.38} & \small{+0.28} & \small{+0.47} & \small{-0.19} & \small{8.02}\\
\small{       } & \small{$\pm$0.01} & \small{$\pm$0.03} & \small{$\pm$0.02} & \small{$\pm$0.01} & \small{$\pm$0.01} & \small{$\pm$0.02} & \small{$\pm$0.01} & \small{$\pm$0.01} & \small{$\pm$0.07} & \small{$\pm$0.07} & \small{$\pm$0.04}\\

\enddata
\label{t1}
\end{deluxetable}

The two groups have:

\begin{center}
[Fe/H]=-1.23$\pm$0.01, [Ba/Fe]= +0.08$\pm$0.02 
\end{center}

and 

\begin{center}
[Fe/H]=-1.22$\pm$0.01, [Ba/Fe]=+0.51$\pm$0.03 
\end{center}

as mean values respectively.
The lower panel of Fig.~\ref{f2} displays the abundances of two other s-elements: Y and Zr.
Again the two groups of Ba-rich and Ba-poor stars are separated, in the sense that
the Y and Zr content is significantly different for Ba-poor and Ba-rich stars. 
The two groups have [Zr/Fe]=+0.14$\pm$0.02, [Y/Fe]=+0.12$\pm$0.02 
and [Zr/Fe]=+0.25$\pm$0.02, [Y/Fe]=+0.23$\pm$0.01 respectively.
The results displayed in the lower panel of Fig.~\ref{f2} can be considered as an independent
confirmation of the presence of two groups of stars with different s-element
content in NGC 1851.

An interesting result comes from Fig.~\ref{f1} (lower panel) which
clearly shows that all Ba-poor stars are located in the bluer RGB, while all
Ba-rich stars are located in the redder one. The two sequences seem to be well separated.
As in the case of M22 \citep{mar09}, our result implies that s-elements
allow us define two distinct populations also in NGC1851.\\

Now, it is instructive to analyze the $\alpha$-element content of the two
groups. We included in this analysis also Ni, another iron-peak element. For this purpose, we plot in
Fig.~\ref{f3} Ni vs. Ba (upper-left panel), Si vs. Ba (upper-right panel), Ca
vs. Ba (lower-left panel), and Ti vs. Ba (lower-right panel).\\
The average value of Ni, Si, Ca, and Ti for the two different Ba groups is summarized in
Table~\ref{t1}, with the corresponding internal dispersion of the mean plotted
in the figure.
As already found for iron (Fig.~\ref{f2}), the Ba-rich and Ba-poor groups have no
appreciable difference in their iron-peak element and $\alpha$-element content, which
have similar values within 1-1.5$\sigma$.\\
This result is in contrast with the hypothesis by \citet{lee09}, and consistent 
with \citet{car10}'s statement that most GCs
do not have an appreciable spread in their iron-peak or $\alpha$-element
content. Our results also imply that in the case of NGC1851 pollution by SNeII
cannot explain the double population.\\

As for the lighter elements, Fig.~\ref{f4} (upper panel) 
confirms that the cluster has a very well defined Na-O anticorrelation, as already shown by 
\citet{yon08}.\\
However, here we have an interesting new result. 
The two Ba groups are partially separated in the Na vs. O plot, with some overlap
at $-0.1<$[O/Fe]$<+0.1$.
Ba-poor stars have an average [Na/Fe]=+0.04$\pm$0.10 and [O/Fe]=+0.09$\pm$0.07, while
Ba-rich stars have [Na/Fe]=+0.47$\pm$0.07, [O/Fe]=-0.19$\pm$0.07. \\
We found that Na is correlated also with C and N. 
The two groups of stars have different mean N contents
([N/Fe]=1.09$\pm$0.08 dex for Ba-poor stars, [N/Fe]=0.73$\pm$0.16 for Ba-rich stars, but very
similar mean C content ([C/Fe]=-0.64$\pm$0.09 dex for Ba-poor stars, [C/Fe]=-0.72$\pm$0.08 for Ba-rich stars). 
However, as in the case of M22, C, N and O do not show a bimodal distribution
(Villanova et al., in preparation).
How C,N,O content is related to the double RGB visible in the (v, v-y) CMD is not clear.
We know that the v band contains many CH and CN features which, according
to \citet{mar08}, can contribute to split the RGB in UV or blue CMDs.
CH bands affect also the Ca narrow-band filter used by \citet{lee09}, 
and this could explain their RGB split without invoking a Ca spread.\\
Any interpretation of the CMD and of the spectroscopic results involving C,N,O
should also take into account the evolutionary stage of our targets.
The investigation of light-element correlations  and the relation between
the double RGB and the C,N,O content is beyond the scope of this
letter and will be discussed in detail in a future paper.

Our final interesting result concerns C+N+O. 
Fig.~\ref{f4} (lower panel) shows that the two groups of 
stars have the same C+N+O content, within  1$\sigma$. In particular, the Ba-poor group have 
log$\epsilon$(CNO)=7.99$\pm$0.02, while the Ba-rich have 
log$\epsilon$(CNO)=8.02$\pm$0.04.\\
About the mean C12/C13 content of the two populations we found
they have the same values ($\sim$5) within the errors. We expected
the Ba-rich stars to have a lower C12/C13 value as the results of the contamination
of the pristine gas they formed from by highly processed ejecta.\\
However we point out that our results about C12/C13 are preliminar.
Also the evolutionary stage of the targets must be taken into account, which
could alter the primordial mean C12/C13 ratio of the two populations making
them to assume a similar value.\\
A better estimation of C12/C13 for each targets and a theoretical treatment 
of the expected difference between the two population will be presented in a
future paper.

Figure~\ref{f4} excludes the possibility that 
the second generation of stars have a significantly enhanced C+N+O content, as
suggested by \citet{cas08}, \citet{ven09}, and \citet{yon09}.
Unfortunatly we do not have stars in common with \citet{yon09} for 
a comparison. Here we note that
\citet{yon09} targets are very bright, close to the RGB-tip. 
It is well possible that some of their stars are indeed AGB
stars. In such a case, the chemical pattern shown by those stars may have been
highly altered by internal nuclear processes.\\
However, as suggested by \citet{cas08}, if the first generation had log$\epsilon$(CNO)=8.00 dex,
then a second generation with double C+N+O content would correspond to log$\epsilon$(CNO)=8.30 dex.
At odd with that, a difference in log$\epsilon$(CNO) of 0.03, as we find, implies that 
the Ba-rich population have about 1.1 times the C+N+O content
of the Ba-poor one, which is consistent with a null difference when we consider the measurement
errors.\\
The dicothomy in Ba and other s-process element content, as well as
the clear spread in the CMD of Fig.~\ref{f1} clearly points towards the presence
of two distinct stellar populations in NGC~1851, as found also by
\citet{mil08} in the SGB.
However, the constancy of the CNO content casts doubts on the scenario
of a second population born from material polluted by first generation
intermediate mass stars. The possibility of a merge of two distinct
clusters, as very recently suggested by \citet{car10b}
should be furher explored.

\acknowledgments
S.V. and D.G. acknowledge the support by BASAL/ FONDAP project.\\
G.P. acknowledges the support by MIUR under the program PRIN2007
(prot. 20075TP5K9).\\
The author acknowledge F. Grundahl ad Y. Momany which kindly provided
the photometry.

\end{document}